\documentclass[reprint, amsmath,amssymb, aps, superscriptaddress, floatfix]{revtex4-1}

\usepackage[utf8]{inputenc}
\usepackage{graphicx}
\usepackage{dcolumn}
\usepackage{bm}
\usepackage{braket}
\usepackage{color}
\usepackage{units}
\usepackage{dsfont}
\usepackage{float}
\usepackage{hyperref}
\usepackage{xcolor}

\begin{document}
\title{Achieving ultimate noise tolerance in quantum communication}

\author{Fr\'ed\'eric Bouchard}
\affiliation{National Research Council of Canada, 100 Sussex Drive, Ottawa, Ontario K1A 0R6, Canada}
\author{Duncan England}
\affiliation{National Research Council of Canada, 100 Sussex Drive, Ottawa, Ontario K1A 0R6, Canada}
\author{Philip J. Bustard}
\affiliation{National Research Council of Canada, 100 Sussex Drive, Ottawa, Ontario K1A 0R6, Canada}
\author{Kate L. Fenwick}
\affiliation{National Research Council of Canada, 100 Sussex Drive, Ottawa, Ontario K1A 0R6, Canada}
\affiliation{Department of Physics, University of Ottawa, Advanced Research Complex, 25 Templeton Street, Ottawa ON Canada, K1N 6N5}
\author{Ebrahim Karimi}
\affiliation{Department of Physics, University of Ottawa, Advanced Research Complex, 25 Templeton Street, Ottawa ON Canada, K1N 6N5}
\affiliation{National Research Council of Canada, 100 Sussex Drive, Ottawa, Ontario K1A 0R6, Canada}
\author{Khabat Heshami}
\affiliation{National Research Council of Canada, 100 Sussex Drive, Ottawa, Ontario K1A 0R6, Canada}
\affiliation{Department of Physics, University of Ottawa, Advanced Research Complex, 25 Templeton Street, Ottawa ON Canada, K1N 6N5}
\author{Benjamin Sussman}
\affiliation{National Research Council of Canada, 100 Sussex Drive, Ottawa, Ontario K1A 0R6, Canada}
\affiliation{Department of Physics, University of Ottawa, Advanced Research Complex, 25 Templeton Street, Ottawa ON Canada, K1N 6N5}

\begin{abstract}
At the fundamental level, quantum communication is ultimately limited by noise. For instance, quantum signals cannot be amplified without the introduction of noise in the amplified states. Furthermore, photon loss reduces the signal-to-noise ratio, accentuating the effect of noise. Thus, most of the efforts in quantum communications have been directed towards overcoming noise to achieve longer communication distances, larger secret key rates, or to operate in noisier environmental conditions. Here, we propose and experimentally demonstrate a platform for quantum communication based on ultrafast optical techniques. In particular, our scheme enables the experimental realization of high-rates and quantum signal filtering approaching a single spectro-temporal mode, resulting in a dramatic reduction in channel noise. By experimentally realizing a 1-ps optically induced temporal gate, we show that ultrafast time filtering can result in an improvement in noise tolerance by a factor of up to 1200 compared to a 2-ns electronic filter enabling daytime quantum key distribution or quantum communication in bright fibers.  
\end{abstract}

\maketitle

\section{Introduction}

Since its inception in 1984~\cite{bennett1984quantum}, quantum key distribution (QKD) has seen a myriad of major conceptual and technological developments. All of these efforts were aimed at achieving quantum communication systems with larger secret key rates~\cite{islam2017provably}, longer achievable communication distances~\cite{yin2016measurement,lucamarini2018overcoming}, or improved security considerations~\cite{lo2012measurement,sasaki2014practical,lo2005decoy}. At the heart of all these requirements, noise represents one of the main limiting factors. Hence, overcoming noise in quantum communication is one of the most crucial requirements for QKD and, more generally, for entanglement distribution~\cite{ecker2019overcoming}, quantum information processing~\cite{cory1998experimental}, and quantum sensing~\cite{escher2011general}. So far, noise has been avoided using dedicated fibers (dark fibers) and limited night-time operation for free-space and satellite-based quantum communication~\cite{buttler1999practical,liao2017satellite}. Provided that the quantum states are encoded in the photon's polarization, all other photonic degrees of freedom, i.e., position, momentum, frequency, and time, can be filtered accordingly. In particular, to achieve optimal noise filtering, the sender, typically referred to as \emph{Alice}, prepares photons encoded in a single spatial, spectral, and temporal (SST) mode. By doing so, the receiver, typically referred to as \emph{Bob}, can avoid detecting noise residing outside of the single SST mode by applying appropriate filters. In theory, single-mode filtering is the ultimate way to decrease noise without altering the signal, consequently increasing the signal-to-noise ratio directly. However, in practice, efficient filtering of single photons to a single SST mode is challenging and has not been demonstrated in the context of quantum communications.

In this paper, we experimentally demonstrate ultimate noise tolerance in quantum communication by realizing an ultrafast temporal filter that allows the detection of quantum signals in a nearly single SST mode. The ultrafast temporal filtering scheme proposed in this work is based on cross-phase modulation via the optical Kerr effect in a single-mode fiber (SMF). A bright pump pulse induces a birefringence in the SMF causing a polarization rotation of the quantum signal. By placing the SMF between crossed polarizers, the pumped SMF acts as an ultrafast switch~\cite{kupchak2017time,kupchak2019terahertz} for terahertz-bandwidth photons. As a result, we demonstrate a tremendous increase in noise tolerance without significantly altering the quantum signal. Our findings are of particular interest for quantum communication systems operating in extremely noisy conditions, e.g., the operation of QKD coexisting with bright classical optical signals~\cite{patel2012coexistence}, the operation of free-space links in daylight conditions~\cite{buttler2000daylight,liao2017long,avesani2019full} or even satellite-based QKD~\cite{liao2017satellite}. For instance, reducing noise by 3 orders of magnitude is sufficient to allow a QKD system to transition from full-moon-clear-night to clear-daytime conditions~\cite{er2005background}. 

Spectral filtering is commonly employed in QKD using interference filters with bandwidths on the order of a few nanometers. Other techniques may achieve narrower bandwidths, but typically present certain limitations. On the other hand, time filtering can also be jointly employed with the help of fast detectors and time-tagging devices, where gating times are limited by the timing jitter of the single-photon detectors, typically on the order of nanoseconds, and other electronic devices involved. With the recent development of superconducting nanowire single photon detectors, a significant improvement in timing jitters can be achieved, i.e. as low as 50~ps~\cite{caloz2018high}. Unfortunately, they require cryogenic cooling (approximately 1~K). Finally, despite time gating being an effective way to limit the amount of noise introduced in the detection event, noise counts may have another undesired effect, i.e., to limit the overall signal detection rate due to detector dead time and detector saturation. Hence, temporal filtering schemes prior to the single-photon detectors can be highly desirable to increase overall performance of QKD in noisy environments.

\begin{figure}[ht]
	\centering
		\includegraphics[width=0.45\textwidth]{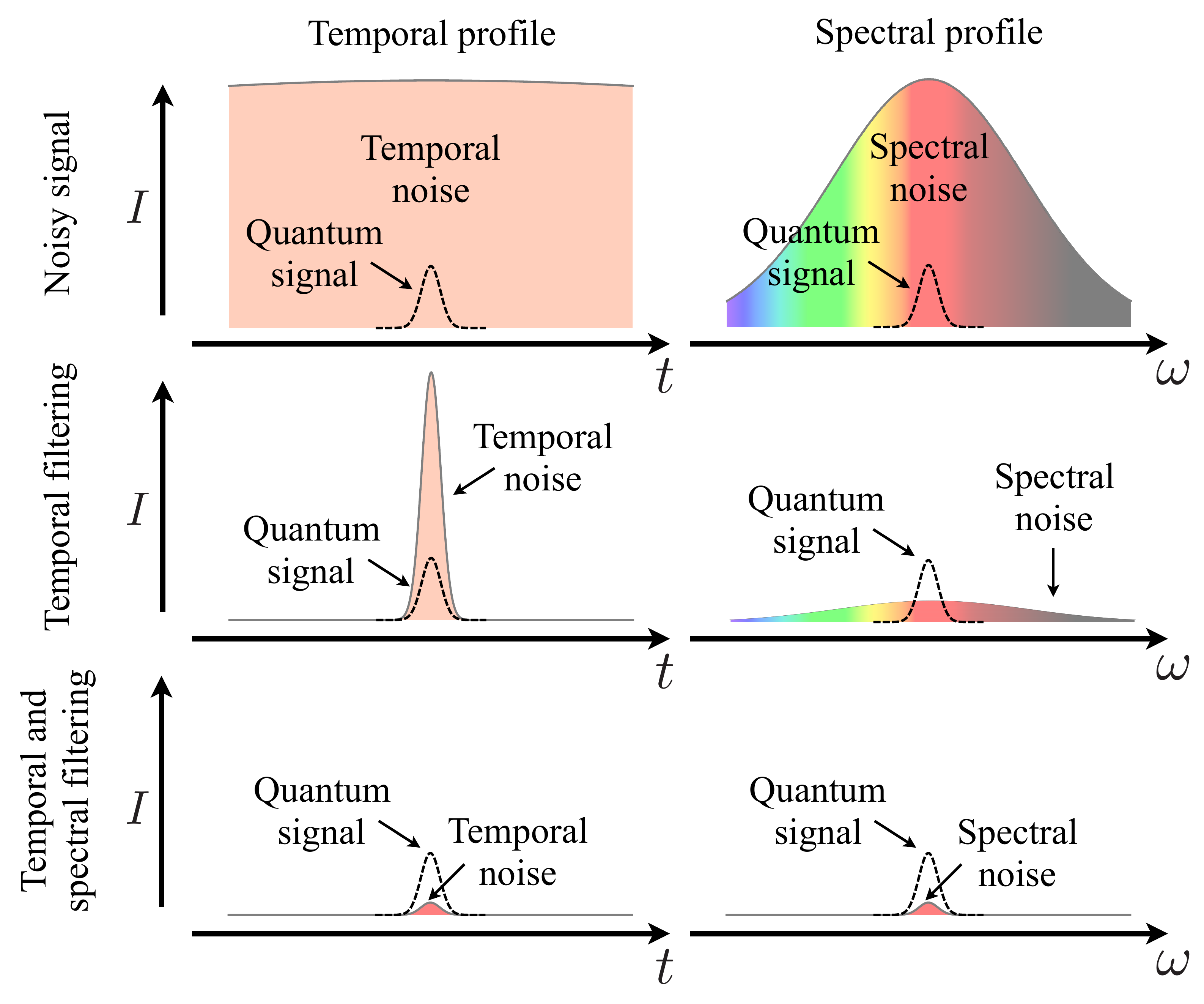}
	\caption{\textbf{Conceptual scheme of ultrafast temporal filtering.} Quantum signals prepared in a single spectrotemporal mode are found in the presence of spectral and temporal noise. As a first step, an ultrafast temporal filter matching the duration of the signal pulse is applied. By doing so, the noise is largely reduced and the quantum signal left untouched. The beam is then sent through a spectral filter that is matched to the bandwidth of the quantum signal, further increasing the signal-to-noise ratio.}
	\label{fig:concept}
\end{figure}

Pulses with minimum time-bandwidth product are known as Fourier-transform-limited (FTL) and are found to occupy a single spectrotemporal mode. For instance, a FTL pulse with a spectral bandwidth of 1~nm at a center wavelength of 800~nm has a pulse duration of approximately 1~ps, which is much lower than the timing jitter of a standard single-photon detector, i.e., approximately 1~ns. Thus, for most QKD experiments reported to date, an improvement factor of up to 1000 can be expected by using an ultrafast temporal filter (UTF). On the other hand, a 1-ns FTL pulse at a center wavelength of 800~nm has a spectral bandwidth of approximately 1~pm, where highly efficient spectral filtering remains challenging~\cite{shan2006free,hockel2009ultranarrow,zentile2015atomic}. By employing the optimal combination of modest spectral filtering and ultrafast temporal filtering, we achieve ultimate noise tolerance; see Fig.~\ref{fig:concept}. 

\begin{figure*}[ht]
	\centering
		\includegraphics[width=0.84\textwidth]{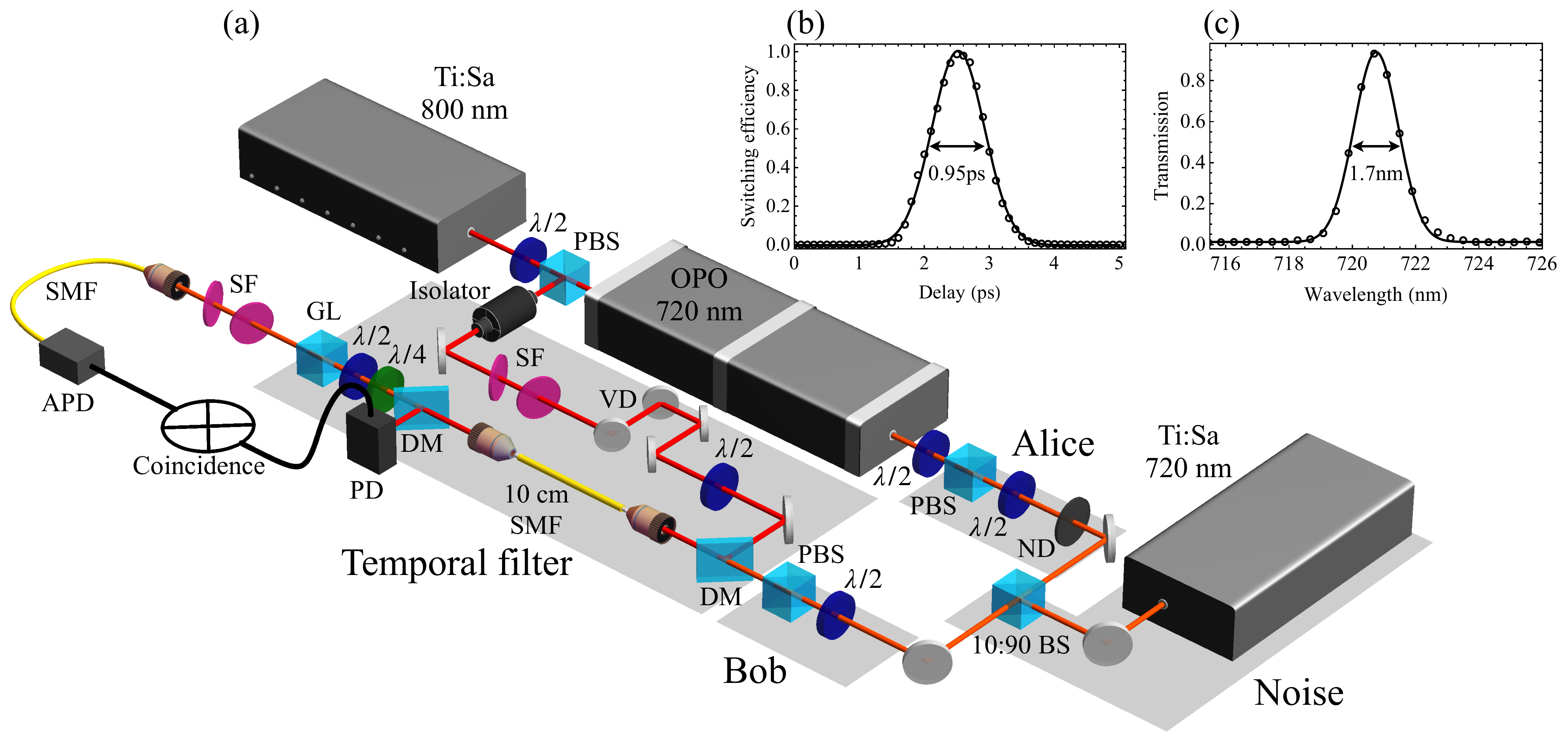}
	\caption{\textbf{Experimental setup.} (a) Experimental setup demonstrating the use of the ultrafast temporal filter in a quantum communication demonstration. Ti:Sa, titanium sapphire laser; $\lambda$/2, half-wave plate; $\lambda$/4, quarter-wave plate; PBS, polarizing beam splitter; OPO, optical parametric oscillator; ND, neutral density filter; BS, beam splitter; DM, dichroic mirror; SF, spectral filters; VD, variable delay; SMF, single-mode fibre; PD, photodiode; GL, glan-laser polarizer; APD, avalanche photodiode detectors. (b) Switching efficiency as a function of the pump delay demonstrating the temporal profile of the UTF. (c) Spectrum of signal photons after spectral filtering at the receiver's stage.}
	\label{fig:setup}
\end{figure*}
Fibre networks will play a fundamental role in the deployment of quantum networks. However, due to their inevitably lossy nature, their reach has been limited to a few hundred kilometers. Thus, in order to complete the quantum network, satellite-based quantum communications and quantum repeaters will bridge the gap to achieve a quantum network by reaching communication distances spanning the entire globe~\cite{kimble2008quantum,simon2017towards,wehner2018quantum}. Very recently, satellite-based quantum communication entered the realm of reality~\cite{vallone2015experimental,liao2017satellite,ren2017ground,liao2018satellite}. In particular, satellite-based quantum communication represents an instance where managing noise is crucial due to the unavoidable detection of background photons from scattered sunlight. Previous experiments of free-space QKD in daylight conditions~\cite{buttler2000daylight,hughes2002practical,peloso2009daylight,liao2017long} demonstrated the feasibility of daylight QKD by applying spatial, spectral, and temporal filters. However, no experiment has so far demonstrated optimal filtering of the spectrotemporal degree of freedom. 
By preparing and measuring the exchanged photons in a single SST mode, we expect a significant improvement in noise tolerance compared to previous results. 

\section{Experiment}

To demonstrate the feasibility of our scheme in quantum communication, we perform a proof-of-principle experiment using our UTF in a polarization-based decoy state BB84 protocol where noise is intentionally introduced in the quantum channel to test the noise tolerance of our technique, see Fig.~\ref{fig:setup}-(a). Weak coherent pulses (WCPs) are prepared by attenuating short pulses at a center wavelength of $\lambda_\mathrm{signal}=720.8$~nm. The WCPs are spectrally filtered to a bandwidth of $\Delta \lambda_\mathrm{signal}=1.7$~nm, corresponding to a transform-limited pulse duration of 0.45~ps. We note that entangled photons, generated for example by spontaneous parametric down-conversion, could also be filtered by the UTF. The WCP are obtained from an optical parametric oscillator pumped by a Ti:sapphire (Ti:Sa) laser with a repetition rate of $f_\mathrm{rep}=80$~MHz. Using a polarizing beam splitter (PBS) and a half-wave plate (HWP), the polarization state of the WCP is prepared at Alice's stage. According to the standard polarization BB84 protocol, Alice randomly prepares the WCP in one of four polarization states, i.e., horizontal, vertical, diagonal, and antidiagonal. The WCP are then attenuated to the single-photon level by using a neutral density filter placed after the polarization encoding stage and a combination of a HWP and a PBS prior to the polarization encoding stage. Using the HWP, the mean photon number of Alice's prepared WCP is set to $\mu$, $\nu$, and 0, manually, corresponding, respectively, to the signal, decoy, and vacuum pulses for the decoy-state protocol~\cite{ma2005practical}. The values of $\mu$ and $\nu$ are varied for different values of channel noise and loss, and subsequently selected offline to optimize the overall secret key rate for each channel condition.

\begin{figure*}[t]
	\centering
		\includegraphics[width=0.95\textwidth]{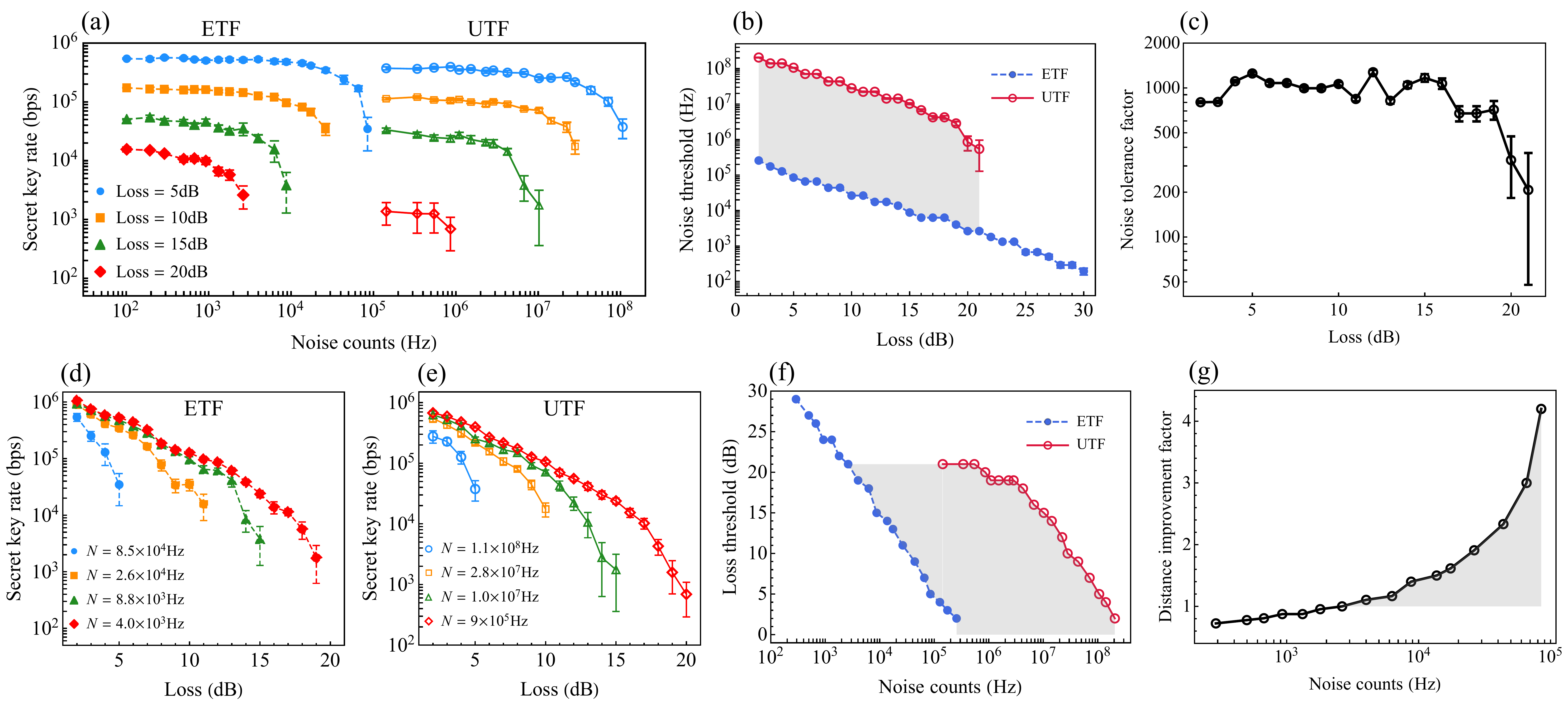}
	\caption{\textbf{Secret key rates}. (a) Secret key rates as a function of noise counts are reported for different channel loss values, $5,10,15,20$~dB. (b) The noise threshold, which is defined as the largest amount of noise for which the mean secret key rate is larger than zero by at least one standard deviation, is shown for different values of channel loss. (c) The noise improvement factor corresponding to the ratio of the noise threshold for the case of UTF over ETF is shown for different values of channel loss. The secret rates as a function of channel loss are reported for (d) ETF and for (e) UTF for different channel noise values, $N$. (f) The loss threshold, which is defined as the largest amount of loss for which the mean secret key rate is larger than zero by at least one standard deviation, is shown for different values of noise counts. (g) The distance improvement factor corresponding to the ratio of the loss threshold for the case of UTF over ETF is shown for different values of noise counts. The shaded areas correspond to channel conditions where an improvement of UTF over ETF is observed.}
	\label{fig:results}
\end{figure*}
The WCPs are then sent through the quantum channel, where a 10:90 (reflection:transmission) beam splitter is used to introduce channel noise. The noise is produced by a second Ti:Sa laser operated in cw mode with a center wavelength and linewidth of ${\lambda_\mathrm{CW}=720.7}$~nm and ${\Delta \lambda_\mathrm{CW}=0.83}$~nm. The amount of noise introduced in the channel is controlled using a combination of HWP, PBS, and neutral density filter. The polarization of the noise source is randomly varied using an additional HWP. The incoming photons are then sent through Bob's detection stage, where the polarization is analyzed using a HWP and PBS, and then through the ultrafast temporal filter. Using a dichroic mirror and a variable delay stage, the signal is made to overlap with a synchronized pump pulse at a center wavelength of ${\lambda_\mathrm{pump}=800~\mathrm{nm}}$, by coupling both into a 10-cm-long SMF with coupling efficiencies of 50~\% and 65~\%, for the signal and the pump, respectively. In the laboratory, femtosecond-scale timing precision is guaranteed by the synchronous nature of the WCP and pump pulse. In remote applications, poor synchronization will result in temporal jitter. The effect of this jitter of the UTF is explored in detail in Appendix F. The switching efficiency, $\eta$, of the quantum signal is given by
\begin{eqnarray}
\eta=\sin^2 (2\theta)\, \sin^2 \left(\frac{\Delta \phi}{2} \right),
\label{eq:switcheff}
\end{eqnarray}
where $\theta$ is the angle between the polarization of the signal and the pump, $\Delta \phi=8 \pi n_2 L_\mathrm{eff} I_\mathrm{pump}/ 3 \lambda_\mathrm{WCP}$ is the nonlinear phase shift induced by the pump in the SMF, $n_2$ is the nonlinear refractive index of the SMF, $L_\mathrm{eff}$ is the effective length of the medium, and $I_\mathrm{pump}$ is the intensity of the pump pulse~\cite{kupchak2019terahertz}. Maximal switching efficiency is observed when $\theta=\pi/4$ and $\Delta \phi=\pi$. Thus, the polarization of the pump is prepared to be at an angle of $45^\circ$ to the polarization of the quantum signal at the input of the SMF. We note that the UTF is positioned after Bob's polarization analysis stage resulting in a fixed signal input polarization state. In order to imprint a uniform nonlinear phase shift across the WCP, we take advantage of the difference in group velocity between the quantum signal and the pump pulse inside the SMF. In particular, we carefully select the pulse duration of the pump and the length of the fiber to allow the pump pulse to fully walk through the WCP within the length of the SMF. This is achieved by spectrally filtering the pump with a pair of angle-tuned bandpass filters such that ${\Delta \lambda_\mathrm{pump}=2.1~\mathrm{nm}}$. Finally, the average power of the pump pulse is set to 300~mW resulting in a unit switching efficiency. 

An avalanche photodiode (APD) is then used to detect the measured photons. The APD is electronically gated using the pump as a reference. The coincidence window is dictated by the timing jitter of the APD and is set to $\Delta t_\mathrm{coinc}=2.0$~ns. By doing so, we achieve a first layer of temporal filtering, which we call electronic temporal filtering (ETF), by reducing continuous noise by a factor of $f_\mathrm{rep} \Delta t_\mathrm{coinc}= 0.16$. Note that for more sophisticated detectors with improved jitter, the coincidence window could be reduced, and the ETF would improve accordingly. Finally, to assess the temporal response of our UTF, we scan the variable delay stage varying the time of arrival of the pump with respect to the signal, see Fig.~\ref{fig:setup}-(b). The switching efficiency at optimal delay is $99\pm1~\%$. The temporal FWHM of the trace is $\Delta t_\mathrm{switch}=0.95\pm0.01$~ps, so is well matched to the duration of the WCPs. To confirm our experimental results, we simulate the temporal profile of our UTF and obtain a good agreement with the experimental results; see Appendix C.

\section{Results}

To assess the feasibility of our temporal filtering scheme in ultrafast quantum communication, we perform a proof-of-principle QKD demonstration, where the figure of merit is given by the secret key rate. In particular, we investigate different channel conditions in terms of noise and loss to demonstrate different regimes where ultrafast quantum communication can offer a considerable advantage over electronic QKD settings. The maximum effective secret key rate $R$ that could be achieved with this apparatus is calculated using the standard decoy BB84 postprocessing procedure~\cite{ma2005practical}. We use the following formula for key generation:
\begin{eqnarray}
R \geq q \left( -Q_\mu f\left(E_\mu \right) H_2\left( E_\mu \right) + Q_1 \left[ 1-H_2\left( e_1 \right) \right] \right),
\end{eqnarray}
where $q=1/2$ is the sifting efficiency, $Q_\mu$ is the gain of signal states, $f(x)$ is the error-correction efficiency, ${H_2(x)=-x \log_2(x)-(1-x) \log_2(1-x)}$ is the binary Shannon entropy function, $E_\mu$ is the quantum bit error rate (QBER), $Q_1$ is the gain of single-photon states, and $e_1$ is the error rate of single-photon states. The experimentally measured gains and QBER, i.e. $Q_\mu$, $Q_\nu$, $E_\mu$, and $E_\nu$, of optimized mean photon numbers $\mu$ and $\nu$ for the signal and decoy states, respectively, are shown in Appendix B for different channel conditions. We use the standard error-correction efficiency factor of $f\left(E_\mu \right)=1.22$.

Experimental results for the secret key rate, $R$, are shown in Fig.~\ref{fig:results}-(a) comparing the case of ETF and UTF as a function of channel noise, $N$, for different values of channel loss. As can be seen, secret key rates can be achieved in a noise regime that is several orders of magnitude larger when operating with UTF (solid curves) compared to ETF (dashed curves). In particular, we compare the noise threshold, i.e., the largest amount of noise for which the mean secret key rate is larger than zero by at least one standard deviation, for both temporal filtering schemes as a function of channel loss, see Fig.~\ref{fig:results}-(b). The noise tolerance factor, i.e., ratio of noise thresholds for the UTF to the ETF, is shown in Fig.~\ref{fig:results}-(c) as a function of channel loss, where a noise improvement factor in excess of 1200 can be obtained, which agrees with values obtained from our simulation; see Appendix C. Our results can also be presented in the context of a second scenario where a fixed amount of noise is present in the quantum channel, but different loss conditions are investigated. The secret key rates are shown as a function of channel loss, see Fig.~\ref{fig:results}-(d) and (e) for various noise counts, $N$. The maximal achievable channel loss can be assessed by considering the loss threshold, see Fig.~\ref{fig:results}-(f). An improvement in communication distance, i.e., distance improvement factor $>1$, occurs for noise counts starting from  $2.6 \times 10^3$~Hz, see Fig.~\ref{fig:results}-(g). Moreover, a maximal distance improvement factor is achieved at a channel noise of $8.5 \times 10^{4}$~Hz with an improvement factor of 4.2.

\section{Discussion and Outlook}

In our experiment, unit switching efficiency is achieved by setting the pump pulse energy to $2.47$~nJ. At this pulse energy, $1.6 \times 10^{-4}$ noise counts per pulse originating from the pump are detected in the single SST mode dedicated to the quantum signal. In particular, the pump generates parasitic nonlinearities such as self-phase modulation and two-photon absorption. These nonlinear processes may create noise photons covering the spectral range of interest for the quantum signals. These noise photons are insignificant in noisy environments where we expect this approach to be most applicable, but become apparent in the low-noise high-loss regime leading to a maximum loss threshold of 21~dB. In Fig.~\ref{fig:results}, the shaded areas represent conditions where QKD cannot work with ETF, but will with UTF. To extend the advantage region of UTF, for instance beyond 21~dB of loss, different avenues can be employed to mitigate the effect of pump noise, e.g., pump pulse and SMF engineering. Hence, the reported pump noise is not a fundamental limitation of our scheme and we expect that lower values of pump noise can be envisaged with further design efforts. Finally, in our experiment, the signal wavelength is set to $\lambda_\mathrm{signal}\sim 720$~nm. This choice is motivated by the available laser source and the desire for low pump noise. Nevertheless, other wavelengths might be of particular interest, e.g., 1310 and 1550~nm. Our proposed UTF scheme can also be applied to the highly desirable telecom wavelengths provided a proper design of the fiber and pump either using our polarization rotation switch or other switching schemes~\cite{hall2011ultrafast}. Finally, let us compare our scheme involving single SST modes to other noise-tolerant QKD schemes. For the case of high-dimensional QKD protocols~\cite{cerf2002security}, the advantage in noise tolerance is observed when a full high-dimensional analysis of multimode quantum signals is carried out compared to a ``coarse-grained" two-dimensional analysis of the multimode signal~\cite{ecker2019overcoming}. Nevertheless, ultimate noise tolerance will be achieved when the signal is encoded and measured in a single SST mode, preventing noise photons from even entering the measurement apparatus in modes other than that used to communicate quantum signals.

In conclusion, we experimentally demonstrate quantum communication with quantum signals in a single SST mode. To show the benefits of our scheme, we have investigate the improvements in terms of noise tolerance and distance improvement for the case of a standard polarization decoy state BB84 QKD protocol for a wide range of channel conditions. We note that our scheme can also be generalized to other degrees of freedom, see Appendix E. In particular, we observe a noise-tolerance improvement factor in excess of 1200 and an improvement in distance by a factor of 4.2 by employing our proposed ultrafast filtering scheme compared to electronic filtering. By doing so, we take advantage of ultrashort pulses in the context of quantum communication. These results can bring daylight free-space quantum communication a step closer to reality. We further expect new features to emanate from joining the fields of ultrafast optics and quantum communication. 

\section{Acknowledgements} 
This work is supported by the National Research Council's High Throughput Secure Networks challenge program, the Joint Centre for Extreme Photonics, Canada Research Chairs, and Canada First Excellence Research Fund. The authors thank Rune Lausten, Denis Guay, and Doug Moffatt for support and insightful discussions.

\section*{Appendix A: Methods}

\begin{figure}[ht]
	\centering
		\includegraphics[width=0.45\textwidth]{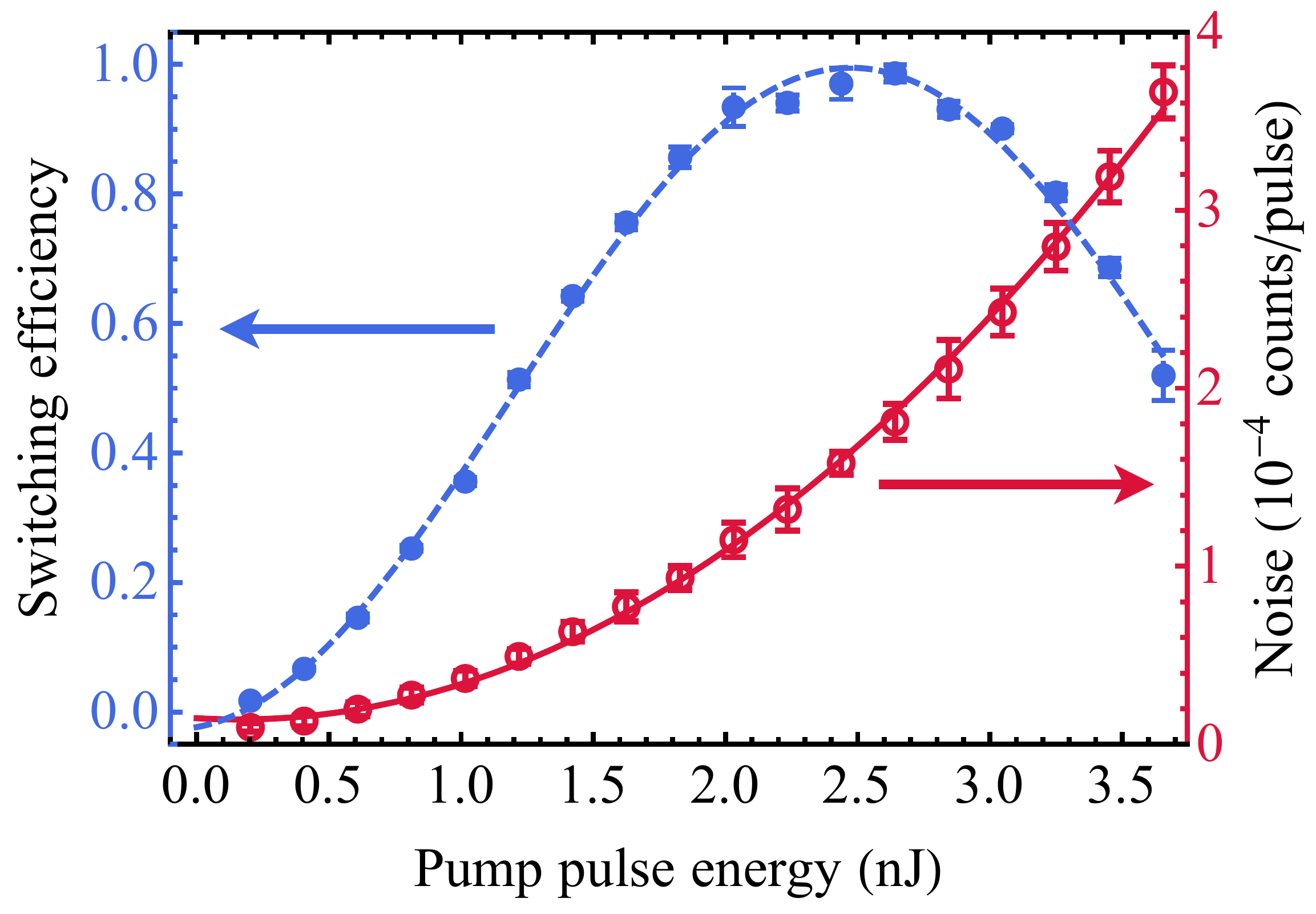}
	\caption{\textbf{Efficiency and noise characteristics of the ultrafast temporal filter}. Switching efficiency (blue dashed curve) and noise counts due to the pump pulse (red solid curve) as a function of the pump pulse energy measured at the output of the SMF. The switching efficiency follows a quadratic sinusoidal relation and the noise curve follows a quadratic relation. The quadratic scaling is typical of third-order nonlinear processes at the origin of the noise counts from the pump.
	}
	\label{fig:switch}
\end{figure}

A Ti:sapphire laser at a center wavelength of $\lambda_\mathrm{pump}=800$~nm is used to pump an optical parametric oscillator (OPO) at a repetition rate of 80~MHz. A signal beam at a center wavelength of $\lambda_\mathrm{signal}=720.8$~nm is achieved through intracavity second-harmonic generation of the OPO signal beam at a center wavelength of $\lambda_\mathrm{OPO}=1441.6$~nm. The pump is further spectrally filtered down to $\Delta \lambda_\mathrm{pump}=2.1$~nm using a pair of tilted interference filters. A second Ti:sapphire laser, serving as the noise source, is operated in cw mode at a center wavelength of ${\lambda_\mathrm{CW}=720.7}$~nm and a linewidth of ${\Delta \lambda_\mathrm{CW}=0.83}$~nm. Noise and loss are controlled using half-wave plates mounted on motorized rotation stages followed with polarizing beam splitters. The total loss of the quantum key distribution channel is varied between 12.3 and 40.3~dB, consisting of the channel attenuation, $t$, varied using a HWP and PBS within a range of 2 to 30~dB. Receiving and detecting losses are given by 3.0-dB coupling loss at the SMF, 0.3-dB loss at the spectral filters, 5.5-dB loss for other optical elements at Bob's stage, and finally 1.5-dB loss due to the efficiency of the APD.
All three beams are coupled to a 10-cm-long single-mode fiber (SMF) (Thorlabs S630-HP, with FC/PC connectors). The pump and the signal beams are coupled into the fiber with a coupling efficiency of 65~\% and 50~\%, respectively. An optical isolator is introduced in the path of the pump pulse prior to the single-mode fiber to limit back reflection of the bright pump pulse into the oscillator. After the SMF, the pump and the signal are separated using a dichroic mirror, where the pump is measured using a fast photodiode to serve as a trigger for the electronic time gating of the quantum signal measured using an avalanche photodiode. A Glan-laser polarizer is then placed in the path of the quantum signal and set to project onto a polarization state that is orthogonal to the polarization of the signal at the input of the SMF, where a HWP and a QWP are used to compensate for polarization rotations induced by the SMF and the DM. A pair of bandpass filters are then used to filter out the signal and noise with a bandwidth of $\Delta \lambda_\mathrm{signal} = 1.7$~nm and a peak transmission efficiency of 93~\%. The two bandpass filters ($>70$\,dB each) and the shortpass filter ($>70$\,dB) provide a total of $>210$\,dB rejection at the pump wavelength. Finally, the quantum signals are then coupled to a SMF with a coupling efficiency of 70~\%, and detected by an APD. The detected signals from the APD are sent into a time-to-digital converter (TDC) (Swabian Instrument, Time Tagger Ultra) for analysis. We note that the TDC has a timing jitter of 22-ps FWHM. Since the TDC has a maximum data-transfer rate of 65~MHz, which is lower than the repetition rate of the pump, we use a conditional filter where all clicks from the fast photodiode that are not followed by a detection event from the APD are discarded directly at the TDC. The coincidence time window is dictated by the timing jitter of the APD and is set to $\Delta t_\mathrm{coinc}=2$~ns. The switching efficiency is measured as a function of pump pulse energy, see blue curve in Fig.~\ref{fig:switch}, following a quadratic sinusoidal relation. A switching efficiency of $99\pm1$~\% is achieved at a pump pulse energy of 2.47~nJ. As a second step, the signal beam is blocked, and the number of noise counts per pulse originating from nonlinear processes in the SMF due to the pump is shown as a function of pump pulse energy, see red curve in Fig.~\ref{fig:switch}. In this case, the noise counts follow a quadratic curve, which is a typical power dependence for third-order nonlinear processes such as two-photon absorption. Finally, we note that in our experiment, the ultrafast temporal filter includes additional loss compared to the electronic temporal filter. Losses due to the optical elements specific to the UTF are removed from the analysis of the ETF data. In particular, this includes loss due to the dichroic mirror combining the pump and signal, the waveplates compensating birefringence in the SMF, the pump spectral filter, the final polarizer, and the coupling efficiency to the final SMF, for a total of 2.05~dB.

\section*{Appendix B: Experimental QKD parameters}

\begin{figure*}[ht]
	\centering
		\includegraphics[width=0.9\textwidth]{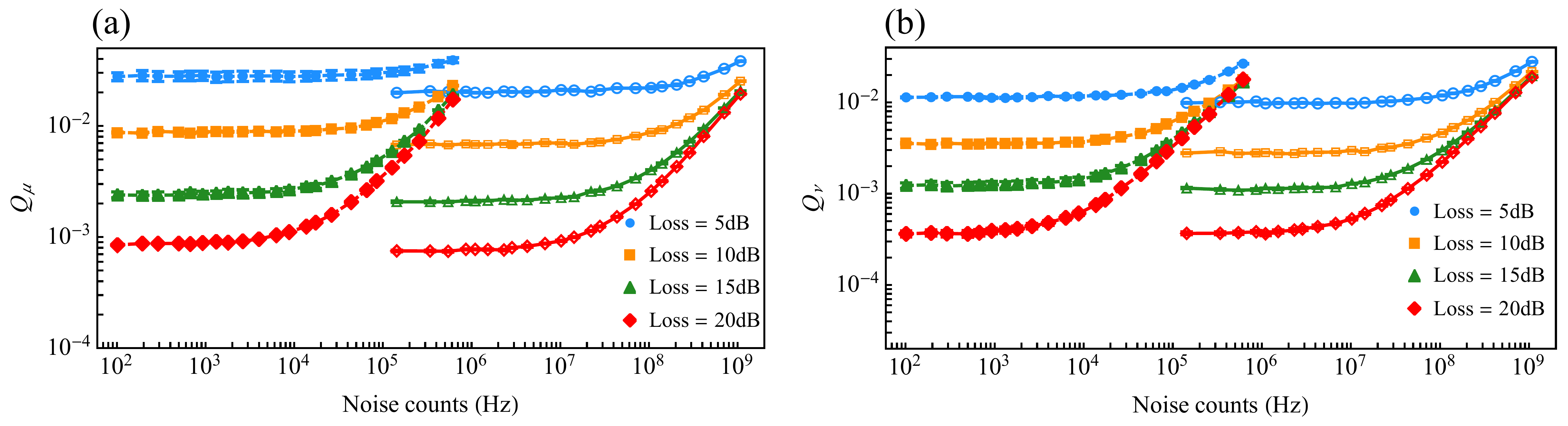}
	\caption{\textbf{Gain}. Gain for (a) signal pulses $\mu$ and (b) decoy pulses $\nu$ as a function of noise counts are reported for different channel loss values, $t=5,10,15,20$~dB, respectively blue circles, orange squares, green triangles and red diamond, for the case of UTF (solid curves and empty markers) and ETF (dashed curves and full markers), respectively on the right and the left. Noise counts correspond to counts detected at the APD within the time window of $\Delta t_\mathrm{coinc}=1$~ns.}
	\label{fig:gain}
\end{figure*}

\begin{figure*}[ht]
	\centering
		\includegraphics[width=0.9\textwidth]{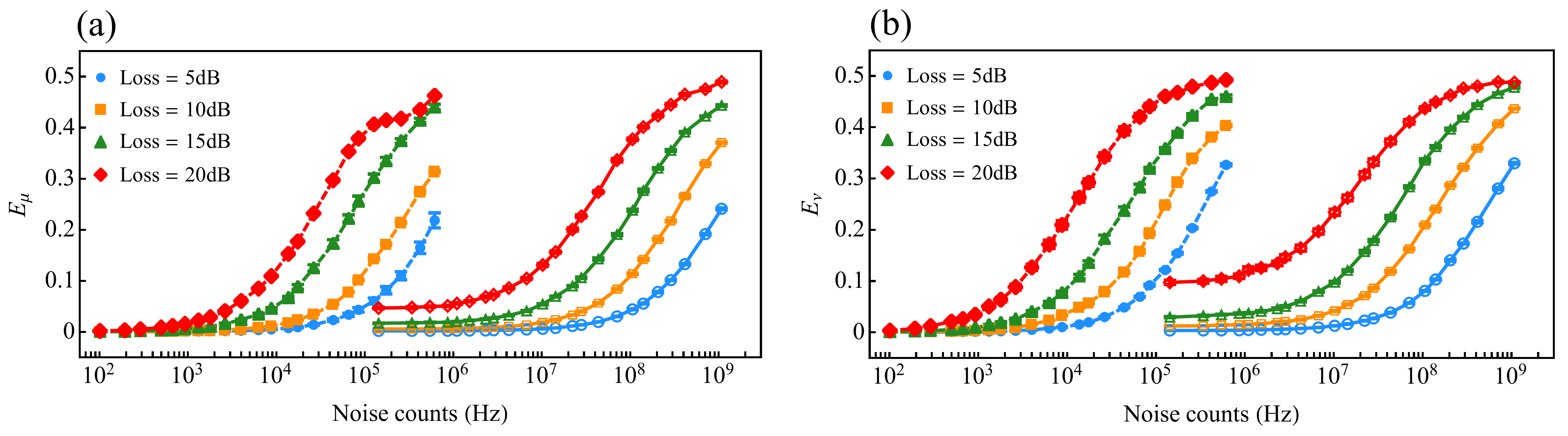}
	\caption{\textbf{Error rate}. Error rate for (a) signal pulses $\mu$ and (b) decoy pulses $\nu$ as a function of noise counts are reported for different channel loss values, $t=5,10,15,20$~dB, respectively, blue circles, orange squares, green triangles, and red diamond, for the case of UTF (solid curves and empty markers) and ETF (dashed curves and full markers), respectively, on the right and the left. Noise counts correspond to counts detected at the APD within the time window of $\Delta t_\mathrm{coinc}=1$~ns.}
	\label{fig:error}
\end{figure*}

The lower bound for the single-photon gain $Q_1$ and the upper bound for the single-photon error rate $e_1$ are, respectively, given by:
\begin{eqnarray}
Q_1 &\geq& \frac{\mu^2 e^{-\mu}}{\mu \nu - \nu^2} \left(Q_\nu e^\nu - Q_\mu e^\mu \frac{\nu^2}{\mu^2} - \frac{\mu^2 - \nu^2}{\mu^2} Y_0 \right),
\end{eqnarray}
\begin{eqnarray}
e_1 \leq \frac{E_\nu Q_\nu e^\nu - Y_0/2}{Q_1 \nu/\left( \mu e^{-\mu} \right)},
\end{eqnarray}
where $Q_1$ is the gain of single-photon states, $e_1$ is the error rate of single-photon states, $Y_0$ is the background rate per pulse, $\mu$ and $\nu$ are the signal- and decoy-state mean photon numbers, $Q_\mu$ and $Q_\nu$ are the signal- and decoy-state gains, and $E_\mu$ and $E_\nu$ are the signal- and decoy-state quantum bit error rates, see Fig.~\ref{fig:gain} and Fig.~\ref{fig:error}. Optimal mean photon numbers $\mu$ and $\nu$ are set to 0.6 and 0.3, respectively.

\begin{figure}[t]
	\centering
		\includegraphics[width=0.48\textwidth]{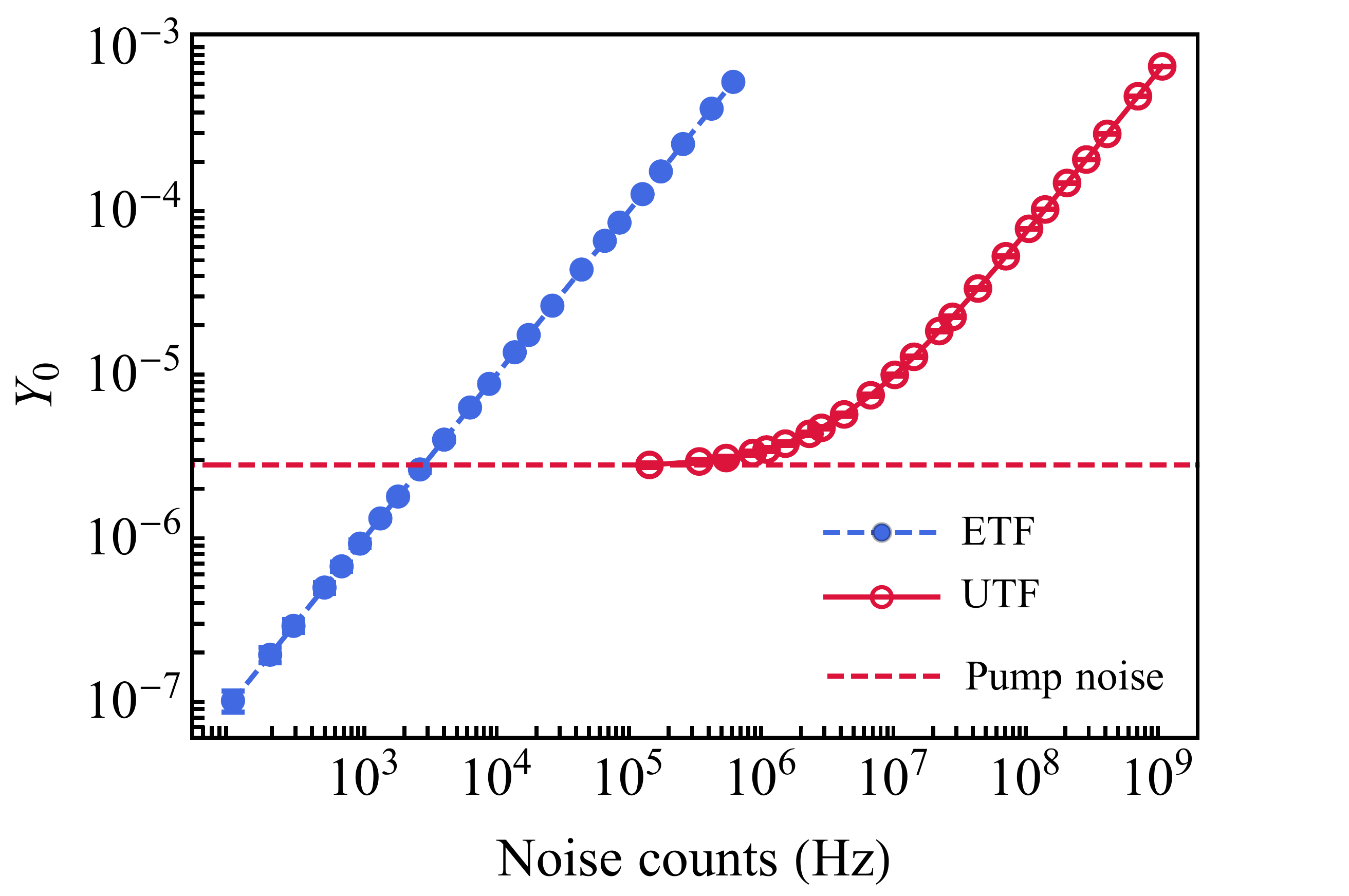}
	\caption{\textbf{Background rate}. Background rate per pulse, $Y_0$, as a function of noise counts that are introduced in the quantum channel by the cw beam. In the case of the UTF, the background rate $Y_0$ is lower bounded by the pump noise counts, i.e., $2.8 \times 10^{-6}$, coming from parasitic nonlinear processes in the SMF.}
	\label{fig:Y0}
\end{figure}

For electronic time filtering, $Y_0$ is mainly a result of the dark counts of the APD, $p_d=100$~Hz, and the noise  introduced in the channel. In the case of the ultrafast time filtering, $Y_0$ is also a result of noise counts from the pump due to nonlinear processes and the noise photons introduced in the channel, see Fig.~\ref{fig:Y0}.

\section*{Appendix C: Simulation of the ultrafast switch}

\begin{figure*}[t]
	\centering
		\includegraphics[width=0.6\textwidth]{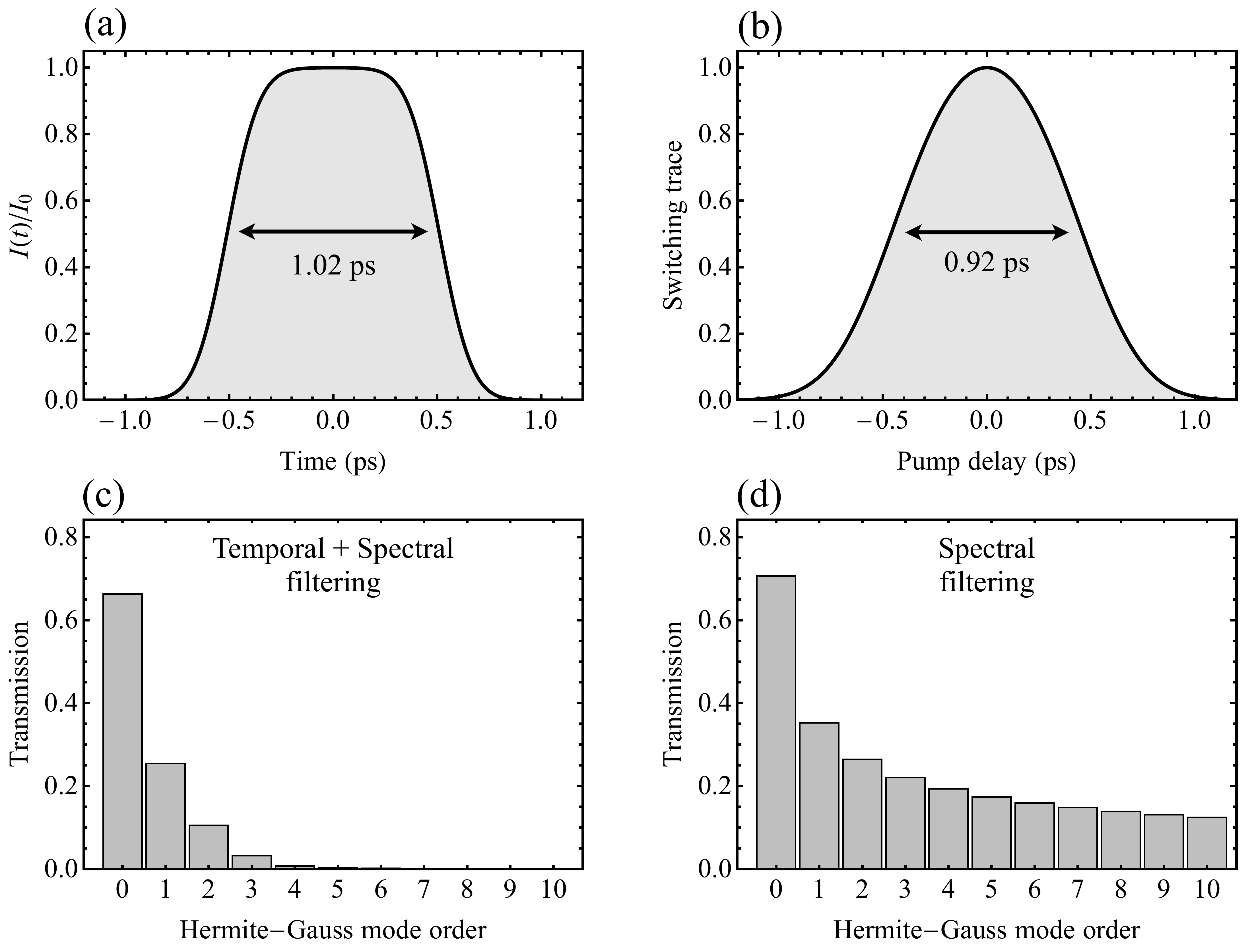}
	\caption{\textbf{Numerical simulation of the ultrafast switch}. (a) Simulated switching response for the following input experimental parameters: pump pulse energy (2.47~nJ), center wavelength ($\lambda_\mathrm{pump}=800$~nm), spectral bandwidth ($\Delta \lambda_\mathrm{pump}=2.1$~nm), signal wavelength ($\lambda_\mathrm{signal}=720.8$~nm), and length of the SMF (10~cm). The switching time given by the FWHM of the switching response is given by 1.02~ps. (b) The switching trace is calculated by calculating the switching efficiency of a signal pulse with a spectral bandwidth of $\Delta \lambda_\mathrm{signal}=1.7$~nm as a function of the delay between the pump and signal pulse. The FWHM of the switching trace is given by 0.92~ps. Numerical evaluation of the transmission of various Hermite-Gauss modes of mode order ranging from 0 to 10 for combined temporal and spectral filter (c) and for only spectral filtering (d).}
	\label{fig:sim}
\end{figure*}

We confirm our experimental results by performing a simulation of the ultrafast temporal switch. The intrinsic switching profile, $\eta$, is given by
\begin{equation}
\eta = \textrm{sin}^2(2\theta)\,\textrm{sin}^2\bigg(\frac{\Delta\phi}{2}\bigg),
\label{eq:efficiency}
\end{equation}
where $\theta$ defines the angle between the pump and signal beam polarizations. Maximal switching occurs when the pump beam polarization is oriented $45^{\circ}$ relative to the signal beam polarization, so we operate under this condition. Considering the group-velocity mismatch between the pump and the signal, the time-dependent nonlinear phase shift, $\Delta\phi(T)$, is then given by
\begin{equation}
\Delta\phi(T) = \frac{8\pi n_2}{3\lambda_{\textrm{signal}}} \int_0^L I_{\textrm{pump}}(T-d_w z)dz,
\label{eq:phase_shift}
\end{equation}
where $n_2$ is the nonlinear refractive index of the fiber, $\lambda_{\textrm{signal}}$ is the wavelength of the signal light, and $z$ is the propagation distance within the fiber of length $L$. The temporal walkoff per unit length experienced by the pump and signal is given by $d_w=v_{gp}^{-1}-v_{gs}^{-1}$, where $v_{gp}$ and $v_{gs}$ are the pump and signal group velocities, respectively. For this reason, we express the intensity profile of the pump beam, $I_{\textrm{pump}}$, in the frame moving with the signal, i.e., $T=t-z/v_{gs}$, where $t$ is time in the laboratory frame. Given a set of experimental parameters, i.e. the pump pulse energy (2.47~nJ), center wavelength ($\lambda_\mathrm{pump}=800$~nm), spectral bandwidth ($\Delta \lambda_\mathrm{pump}=2.1$~nm), the signal wavelength ($\lambda_\mathrm{signal}=720.8$~nm), and the length of the SMF (10~cm), we can simulate the switch efficiency as a function of time in the co-moving signal frame. From this we can calculate the expected noise improvement factor; see Fig.~\ref{fig:sim}-(a). The switch efficiency shape is set by the pump pulse duration and the temporal walk-off of the pump and signal through 10~cm of SMF, i.e. approximately 1~ps. It is the temporal walkoff between the signal and the pump that explains the top-hat shape of the switch efficiency as a function of time in the co-moving signal frame in Fig.~\ref{fig:sim}~(a). Hence, the temporal walkoff results in a uniform phase shift across the signal pulse. In order to compare our model with our experimental result, we calculate the switched efficiency as a function of pump-pulse-to-signal-pulse delay for a signal pulse at a center wavelength of $\lambda_\mathrm{signal}=720.8$~nm and spectral bandwidth of $\Delta \lambda_\mathrm{signal}=1.7$~nm. The shape of the calculated switched efficiency profile is similar to that from the main text, with a FWHM of 0.92~ps close to the measured value of 0.95~ps.

By integrating the switched efficiency profile, we estimate a noise reduction factor of 1960 compared to an electronic temporal filtering of 2~ns. The discrepancy between this value and the noise improvement factor of up to 1200 reported experimentally in the main text can be explained by the spectral profile of the cw noise photons and the extra loss removed from the analysis of the ETF. In our experiment, a cw source with finite linewidth is used as the source of noise. In particular, the measured noise linewidth is given by $\Delta \lambda_\mathrm{noise}=0.83$~nm, whereas the spectral filtering is given by $\Delta \lambda_\mathrm{filt}=1.7$~nm. This spectral mode mismatch will lead to a noise improvement factor in itself since the considered noise did not perfectly match the single temporal mode considered for the signal. This effect can be seen by measuring the output spectrum of the cw noise after passing through the ultrafast switch, i.e. $\Delta \lambda_\mathrm{Switched CW}=1.54$~nm. By considering a cw beam with a narrow linewidth, i.e., $<<\Delta \lambda_\mathrm{signal}$, a noise improvement factor of 2360 is expected, matching our experimental results.

\section*{Appendix D: Measuring a single spectrotemporal mode}

The benefit of our technique lies in the fact that single spatial and spectrotemporal modes are employed to encode and decode information. By doing so, noise can only enter our measurement apparatus through that single optical mode. However, in our experiment we do not exactly perform filtering of a single spectrotemporal mode. Indeed, by using a sequence of filtering in the temporal then spectral domain, we can achieve a filtering down to a very small number of optical modes, see Fig.~\ref{fig:sim}-(c),(d). However, even though a larger signal-to-noise ratio would be achieved by projecting on a single spectrotemporal mode, the currently available techniques to do so still possess modest efficiencies. Thus by sequentially filtering both temporally and spectrally, we can achieve ultimate noise tolerance by adjusting the level of filtering to optimize both the signal-to-noise ratio as well as the total efficiency. In order to assess the single-mode character of our filtering technique, we numerically calculate the transmission of different spectrotemporal modes through the sequence of ultrafast temporal and spectral filtering, see Fig.~\ref{fig:sim}-(c), compared to only spectral filtering, see Fig.~\ref{fig:sim}-(d). For Hermite-Gauss modes with mode order larger than 4, the transmission for the case of temporal and spectral filter is already lower than 0.7~\%. However, for spectral filtering alone, we see a significant contribution to noise due to higher-order modes.

\section*{Appendix E: Implementation to other degrees of freedom}

We demonstrate the performance of our UTF scheme in a polarization-based QKD proof-of-principle experiment. However, our technique is not limited to this specific degree of freedom, i.e. polarization. Other degrees of freedom, such as time bins, phase coding, and spatial modes, can be advantageous for various experimental implementations. Hence, achieving ultimate noise resistance in quantum communication systems employing these various degrees of freedom can be critical. Here, we briefly describe how our UTF technique can be generalized to other degrees of freedom.

\subsection*{Time bins}

A common degree of freedom employed in QKD is the temporal degree of freedom with time-bin encoding. Typically, a signal is prepared as a train of pulses separated by a time determined by the temporal resolution of the detectors (timing jitter) or the speed of the source modulators. The information can either be encoded in the time of arrival (computational basis) or by the relative phase between pulses, which is known as phase coding and will be described in the following section. In the computational basis, time-bin encoding is advantageous since it can propagate through a fiber channel with minimal disturbance and can be straightforwardly detected with a single-photon detector. Moreover, the number of time bins can be extended beyond the standard two-dimensional configuration where larger information capacity could be achieved and the UTF could still be employed. However, for noise considerations, lower dimensions result in larger noise tolerance. Hence, the time separation between the two time-bins may be selected at will to match experimental requirements, e.g., nanoseconds time separation, but the time bin pulses themselves must be in near single spectrotemporal modes, as described in the main text. By doing so, only two optical modes are involved in the generation and detection of the signals dramatically reducing the effect of noise. 

The ultrafast temporal filter can then be applied to the time-bin state by shaping the pump pulse to match the pulse shape of the signal, i.e. a train of pump pulses aligned with the train of signal pulses. The pump will then switch only the two optical modes occupied by the signal. After going through the UTF, the signal is detected by a single-photon detector where the time of arrival of the signal is recorded and used to generate the raw key in the QKD protocol.

\subsection*{Phase coding}

A fundamental concept in QKD is the necessity to encode and decode information in conjugate bases. If time-bin encoding represents the temporal computational basis, the conjugate bases are achieved via phase coding where information is encoded in the relative phase of the signal pulses. There exist various phase-coding protocols, e.g., differential phase shift, coherent one-way, and round-robin differential phase shift. In all these protocols, the phase is ultimately measured using some sort of interferometric apparatus. In the simplest case, the relative phase between two time bins can be measured using an unbalanced Mach-Zehnder interferometer and two single-photon detectors, i.e. one at each output port of the interferometer. The relative phase between the pulses will dictate whether constructive or destructive interference will take place at the output beam splitter resulting in a detection event in the first or the second output port, respectively. 

As in the case of time-bin encoding, the signal must be generated in near-single spectrotemporal mode pulses at different arrival times. The UTF can then be introduced before the interferometer to filter out noise outside the optical modes considered. Since the UTF acts only on the intensity of the signal states and does not alter their phase, the phase information between the time bins is left undisturbed by the UTF. 

\begin{figure*}[t]
	\centering
		\includegraphics[width=0.7\textwidth]{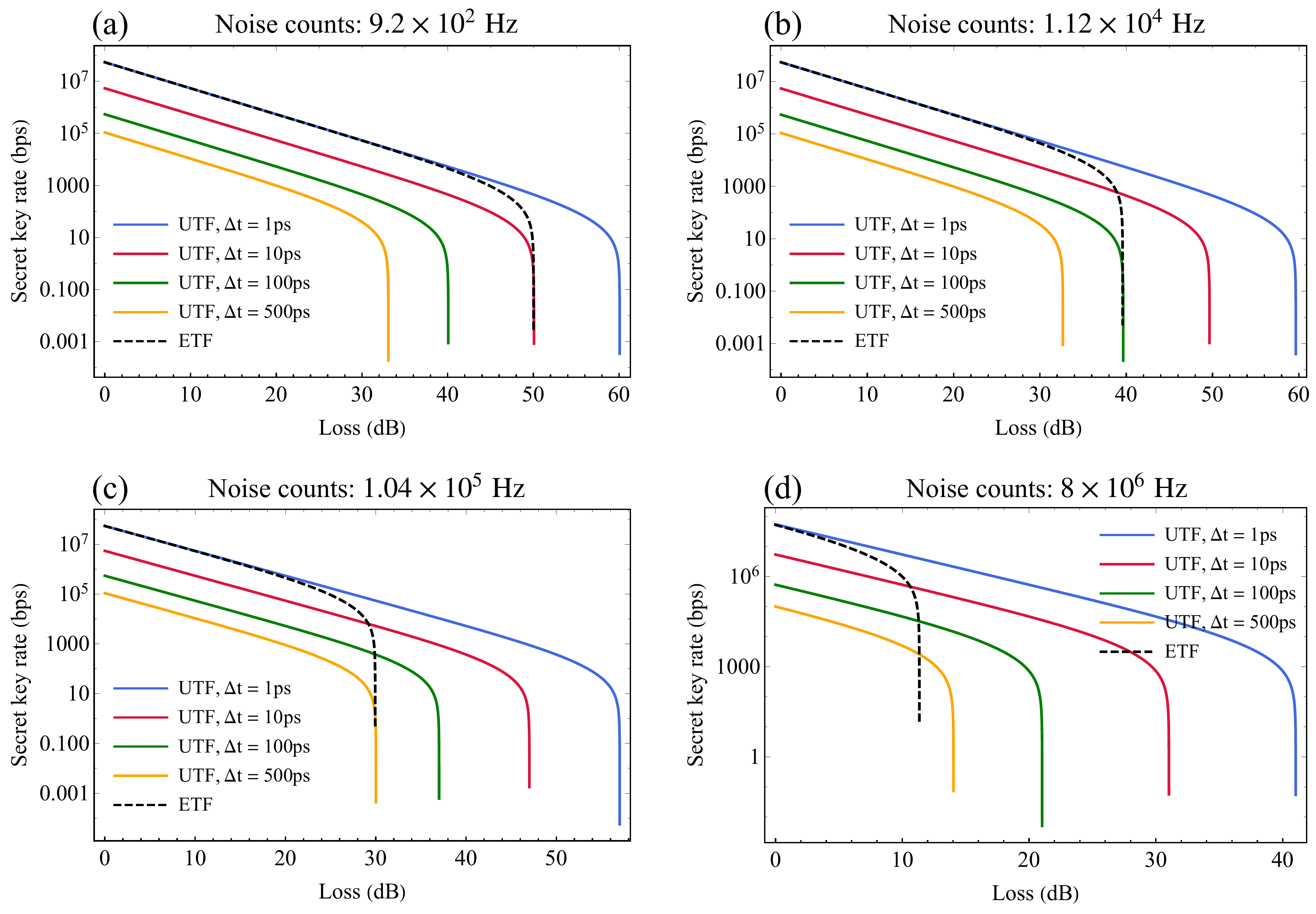}
	\caption{\textbf{Numerical simulation of the ultrafast switch in the presence of temporal fluctuations}.}
	\label{fig:tempfluc}
\end{figure*}

\subsection*{Frequency bins}

Similar to time bins, information can be encoded and decoded in frequency bins. Encoding in the frequency degree of freedom can still be achieved in the presence of our UTF with a near-unity efficiency. In practice, the intensity of the pump is adjusted to achieve a 100~\% switching efficiency of the quantum signal at a certain wavelength. For the case of a dual-wavelength implementation a few nanometers apart, switching efficiency in excess of 99~\% can still be achieved over the various wavelengths.

\subsection*{Spatial modes}

Finally, another degree of freedom that has received attention for implementation in a QKD system are transverse spatial modes. In particular, orbital angular momentum (OAM) states of light have been demonstrated in QKD implementations over free-space channels. At first glance, our UTF may not seem compatible with higher-order spatial modes since it requires coupling the signal into a single-mode fiber. However, the UTF can simply be introduced after the measurement of the spatial modes.

Indeed, most spatial mode techniques are based on the fact that a single mode fiber can act as a spatial-mode filter for the fundamental Gaussian mode. Thus, if a higher-order spatial mode can be turned into the fundamental Gaussian via phase manipulation (using a spatial light modulator or refractive elements) then coupled to a single-mode fiber, that higher-order mode can be measured. Thus, since the last step of a spatial mode measurement scheme is coupling to a single mode fiber, the UTF can be introduced at that stage with high efficiency due to the spatial mode matching.

\section*{Appendix F: Ultrafast temporal filtering in the presence of temporal fluctuations}
In our proposed scheme, fluctuations in the time of arrival and polarization of the pulses from the sender to the receiver can result in a serious limitation for practical implementations. Indeed, temporal fluctuations can arise from different causes, e.g., atmospheric turbulence, pulse broadening, or low synchronization precision. Although not a fundamental restriction, temporal fluctuations may hinder the practicality and efficiency of our scheme using UTF for quantum communication in noisy conditions. We investigate this issue by performing a simulation of a noisy QKD channel in the presence of temporal fluctuations.

In our simulation, we consider that Alice is preparing pulses that are 1~ps in duration and that Bob applies a 1-ps temporal filter. Four different scenarios are considered, namely temporal fluctuations from the channel resulting in pulse duration $\Delta t$ of 1, 10, 100, and 500~ps at the receiver. As a comparison, we consider the case of electronic time filtering with an electronic gating time of 1~ns where temporal fluctuations below 1~ns will have no effect on the secret key rate. Moreover, we consider in our simulation a polarization visibility of 0.99, a detector efficiency of 0.8, detector dark counts of 100~Hz, a repetition rate of 80~MHz, and for simplicity we are assuming single-photon pulses. 

In particular, we evaluate four different noise scenarios, see Fig.~\ref{fig:tempfluc}. In the first case, see Fig.~\ref{fig:tempfluc}-(a), a noise count rate of $9.2\times 10^{2}$~Hz is considered, where the loss threshold of ETF is similar to the loss threshold of UTF with temporal fluctuations $\Delta t = 10$~ps. This means that for this particular level of noise, if the temporal fluctuations are larger than $\Delta t=10$~ps, the ETF outperforms the use of UTF. However, for larger noise levels, e.g., a noise count rate of $8 \times 10^6$~Hz, see Fig.~\ref{fig:tempfluc}-(d), the use of UTF even in the presence of temporal fluctuations with $\Delta t=500$~ps, a larger loss threshold is observed. Intermediate noise levels are also simulated.

In summary, temporal fluctuations can affect the performance of QKD in the presence of UTF. Nevertheless, for large enough levels of noise, as is the case in daylight conditions, UTF can still be beneficial even in the presence of temporal fluctuations. 


%

\end{document}